\documentclass[conference]{IEEEtran}

\IEEEoverridecommandlockouts
\usepackage{cite}
\usepackage{amsmath,amssymb,amsfonts}
\usepackage{algorithmic}
\usepackage{graphicx}
\usepackage{textcomp}
\usepackage{xcolor}
\usepackage{booktabs}
\def\BibTeX{{\rm B\kern-.05em{\sc i\kern-.025em b}\kern-.08em
    T\kern-.1667em\lower.7ex\hbox{E}\kern-.125emX}}
\begin{document}

\title{LR-FHSS-Sim: A Discrete-Event Simulator for LR-FHSS Networks}

\author{\IEEEauthorblockN{Jean Michel de Souza Sant'Ana, Arliones Hoeller Jr., Hirley Alves, Richard Demo Souza}
\IEEEauthorblockA{\textit{Centre for Wireless Communications, University of Oulu}, Oulu, Finland\\
\textit{Telecommunications Engineering Department, Federal Institute of Santa Catarina}, São José, Brazil\\
\textit{Electrical \& Electronics Engineering Department, Federal University of Santa Catarina}, Florianópolis, Brazil\\
\{jean.desouzasantana, hirley.alves\}@oulu.fi, arliones.hoeller@ifsc.edu.br, 
richard.demo@ufsc.br}}

\maketitle

\begin{abstract}
This work presents the LR-FHSS-Sim, a free and open-source discrete-event simulator for LR-FHSS networks. We highlight the importance of network modeling for IoT coverage, especially when it is needed to capture dynamic network behaviors. Written in Python, we present the LR-FHSS-Sim main structure, procedures, and extensions. We discuss the importance of a modular code, which facilitates the creation of algorithmic strategies and signal-processing techniques for LR-FHSS networks. Moreover, we showcase how to achieve results when considering different packet generation traffic patterns and with a previously published extension. Finally, we discuss our thoughts on future implementations and what can be achieved with them.
\end{abstract}


\section{Introduction}
Non-terrestrial networks (NTN) can enable global access to the Internet of Things (IoT) services regardless of geographical location or terrain. Services in remote areas like transportation, fleet management, logistics, solar, oil and gas extraction, offshore monitoring, utilities smart metering, farming, environment monitoring, and mining are some examples~\cite{Centenaro:CST:2021}. Using Low Power Wide Area Networks (LPWANs) combined with satellite technologies is a promising solution to integrate terrestrial networks with NTNs~\cite{ullah:WCM:2021}.
A promising technology for satellite communication for IoT applications is the Long Range Frequency Hopping Spread Spectrum (LR-FHSS)~\cite{fraire.headerless.23,alvarez:ACS:2022}. As part of the Long Range Wide Area Network (LoRaWAN) specification~\cite{LoRaWAN_Region} but focused on satellite communication, this technique divides the payload into small pieces and transmits it across different physical channels. On top of that, it also transmits several redundant copies of the header into different frequency bandwidth channels.

Research on LR-FHSS shows us that modeling such networks can be quite challenging, where one can make several assumptions to make a simpler model~\cite{ullah.WCL.22}, or work with very detailed models~\cite{maleki.CL.23} even considering the channel effects due to satellite movement. Additionally, the adoption of ``smart'' signal processing and resource allocation techniques is increasing as 6G research moves toward AI integration~\cite{Letaief:JSAC:2022}. All of this can be impractical to describe with relatively simple mathematical frameworks~\cite{Santana:IoTJ:2024}. The use of simulations is quite common, mostly based on discrete Monte Carlo techniques, usually tailor-made to the problem being studied~\cite{alvarez:ACS:2022, ullah:WCM:2021}. This raises some issues like low re-usability and low flexibility for future research.

An alternative approach is the use of discrete-event simulations. The main idea of this technique is to model a system whose global state changes over time. This is relevant, for example, in cases where there is a strong temporal correlation (e.g., sequential transmissions, different transmission durations), especially because they are not easily modeled by relatively simple mathematical frameworks or Monte Carlo simulations. Moreover, in discrete-event simulations, isolating and analyzing parts of a system and changing specific network functionalities without compromising or even changing the rest of the setup becomes easier. Frameworks like ns-3~\cite{ns3} and OMNeT++~\cite{omnet}, for example, implement robust discrete-event simulations for several different technologies and environments, from traditional TCP/UDP internet to complex wireless network environments. Alternatively, Python presents a package called SimPy~\cite{simpy}, which allows the creation of customized simulation environments.

The development and use of discrete-event simulators for LPWAN, particularly for the Chirp Spread Spectrum (CSS) LoRa modulation~\cite{LoRaWAN_spec}, is very common in the literature. We can cite at least four different modules developed for ns-3~\cite{Magrin:ICC:2017,Reynders:ICC:2017, Van:IoTJ:2017, Reynders:IoTJ:2018, To:ICC:2018, Capuzzo:MHN:2018, Haxhibeqiri:IoTJ:2019, Magrin:IoTJ:2020, Finnegan:IoTJ:2020}\footnote{https://github.com/signetlabdei/lorawan}~\footnote{https://github.com/drakkar-lig/lora-ns3-module}~\footnote{https://github.com/imec-idlab/ns-3-dev-git/tree/lorawan}, the FLoRa (Framework for LoRa) for the OMNeT++~\cite{Slabicki:NOMS:2018, Leonardi:IoTJ:2019, Premsankar:TII:2020, Toro:INFOCOM:2021, Jesus:IoTJ:2021}\footnote{https://flora.aalto.fi} and the LoRaSim~\cite{Bor:MSWiM:2016, Voigt:EWSN:2017}\footnote{https://mcbor.github.io/lorasim/} built with the SimPy framework for Python. From the listed simulators, FLoRa and the LoRaWAN ns-3 module~\cite{Magrin:ICC:2017} are the most organized and structured, from which we may attribute their success. However, the modules and simulation setups with them are not centralized, while the original repository still contains only the basic modules. Although LoRaSim was not officially published on a repository and its last update was in 2018, it was the most used and extended LoRa simulator~\cite{Pop:GLOBECOM:2017,Farooq:LCN:2018,Sugianto:EECSI:2018,Ferreira:MobCld:2019,Lee:ICAIIC:2020,Cui:JCN:2020,Lalle:ACS:2021,Charles:IECON:2021,Francisco:ConfTELE:2021,Wongwatthanaroek:JCSSE:2021,Wang:OJIES:2023}. We can refer to its popularity partially due to being the first simulator openly available for LoRa networks. However, even more recent works still considered using it, highlighting that its simplicity may be a decisive factor in such a choice.

\begin{table}[tb]
\centering
\caption{List of representative LoRa publications using discrete-event simulators grouped in different categories.}
\label{tab:references}
\begin{tabular}{@{}lc@{}}
\toprule
\textbf{Category}                    & \textbf{Publications}                           \\ \midrule
Performance and scalability analysis & \begin{tabular}[c]{@{}c@{}}\cite{Magrin:ICC:2017}, \cite{Van:IoTJ:2017}, \cite{Magrin:IoTJ:2020}, \cite{Toro:INFOCOM:2021}, \cite{Bor:MSWiM:2016},\\ \cite{Pop:GLOBECOM:2017}, \cite{Sugianto:EECSI:2018}, \cite{Ferreira:MobCld:2019}, \cite{Francisco:ConfTELE:2021}\end{tabular} \\ \midrule
Access and scheduling algorithms     & \begin{tabular}[c]{@{}c@{}}\cite{Reynders:IoTJ:2018}, \cite{To:ICC:2018}, \cite{Capuzzo:MHN:2018}, \cite{Haxhibeqiri:IoTJ:2019},  \cite{Leonardi:IoTJ:2019},\\  \cite{Lee:ICAIIC:2020}, \cite{Cui:JCN:2020}, \cite{Lalle:ACS:2021}, \cite{Wongwatthanaroek:JCSSE:2021}\end{tabular} \\ \midrule
Parameters optimization          & \begin{tabular}[c]{@{}c@{}}\cite{Reynders:ICC:2017}, \cite{Finnegan:IoTJ:2020}, \cite{Slabicki:NOMS:2018}, \cite{Premsankar:TII:2020},\\ \cite{Jesus:IoTJ:2021}, \cite{Charles:IECON:2021}, \cite{Wang:OJIES:2023}\end{tabular} \\ \midrule
Multiple gateways analysis           &  \cite{Voigt:EWSN:2017}, \cite{Farooq:LCN:2018}     \\ \bottomrule
\end{tabular}
\end{table}

We list in Table~\ref{tab:references} what we consider to be the main categories of works utilizing discrete-event simulators for LoRa networks, along with a list of some representative papers in each category. As expected, several works on performance and scalability analysis are used since they tend to be the first step when validating a simulator. However, we can see many works on scheduling algorithms, optimizations, and even multiple gateways. These techniques can be very challenging to model only mathematically. For example, scheduling and online parameter optimization algorithms tend to evolve with time until they reach a certain stability. Ultimately, in the case of dynamic networks, mathematical modeling might become of utmost complexity. Finally, the analysis of multiple-gateway scenarios is very dependent on the position of the gateways; thus, closed-form equations are impractical in most cases.

\subsection{Contributions}
Drawing inspiration from LoRaSim for CSS LoRa, this work introduces an open-source LR-FHSS discrete-event simulator coded in Python utilizing the SimPy framework. Unlike prior LR-FHSS works, our simulation environment is not tailored to specific scenarios and can accommodate various algorithms and signal processing techniques. Unlike LoRaSim, we have enhanced the code structure and publication methods to foster usability across diverse research projects. Unlike FLoRa, LR-FHSS-Sim is designed to be modular, allowing researchers to employ the necessary structures for simulations selectively and develop new modules with additional functionalities.

\subsection{Organization}
This paper is organized as follows. In Section~\ref{sec:sim}, we present a brief description of the LR-FHSS technology, the simulator architecture, and an example of an extension module. Section~\ref{sec:results} presents some results with the simulator for a baseline model, reproducing one published result using one extension and an example of a novel extension module. Finally, Section~\ref{sec:conclusions} concludes the paper.

\section{LR-FHSS-Sim}\label{sec:sim}
This section presents the LR-FHSS-Sim (LoRa - Frequency Hopping Spread Spectrum Simulator). The simulator is available freely on its online repository\footnote{https://github.com/Xexell/LR-FHSS-sim}. However, before diving into it, we discuss some basic concepts of LR-FHSS.

LR-FHSS is a Frequency-Hopping Spread Spectrum technique, part of the LoRaWAN specification, designed mainly for uplink IoT direct-to-satellite communications~\cite{boquet.CM.21}. An LR-FHSS transmission hops over multiple frequencies (the so-called physical channels) while employing GMSK modulation at a bandwidth of $488$~Hz. The total bandwidth of the LR-FHSS channel and the minimum physical channel separation of the LR-FHSS hopping depend on regional specifications~\cite{LoRaWAN_Region}. This results in a different number of physical channels of $488$~Hz available for use during one single transmission.
For example, in Europe, there are 8 options of 35 or 86 physical channels to hop, depending on the chosen data rate. These options are the so-called channels' grids, illustrated in~\cite{fraire.headerless.23,Santana:IoTJ:2024}.

The LR-FHSS packet structure comprises a header and a payload~\cite{LoRaWAN_Region}. The header contains metadata about the transmission parameters and, most importantly, the seed to generate the pseudorandom hop sequence defining the physical channels assigned to the packet elements during that transmission. Usually, the end device sends two or three header copies of $t_{header} = 233.472$~ms sequentially across different physical channels. The payload is channel coded according to a convolutional code of coding rate (CR) of 1/3 or 2/3 and split into small fragments of duration $t_{f} = 102.4$~ms~\cite{LoRaWAN_Region}. The number of payload fragments is given by~\cite{Santana:IoTJ:2024}
\begin{align}
    f = \left\lceil \frac{b+3}{6~\textsf{CR}} \right\rceil,
\end{align}
where $b$ is the payload length (in bytes) and $\textsf{CR} \in\{1/3,2/3\}$ represents the coding rate.

Data transmission follows the unslotted ALOHA multiple access protocol, where end devices randomly pick one seed and generate a pseudorandom sequence of physical channels. During transmission, the end device hops to a new physical channel for every packet element (headers and fragments). A receiver that successfully decodes a header can recover the pseudorandom sequence and acquire knowledge of the physical channels used for the transmission. Finally, a packet is successfully decoded if at least one header and 1/3 or 2/3 of the total fragments are received, depending on the employed coding rate.

\subsection{Architecture}

\begin{figure*}[t]
    \centering
    \includegraphics[width=0.8\linewidth]{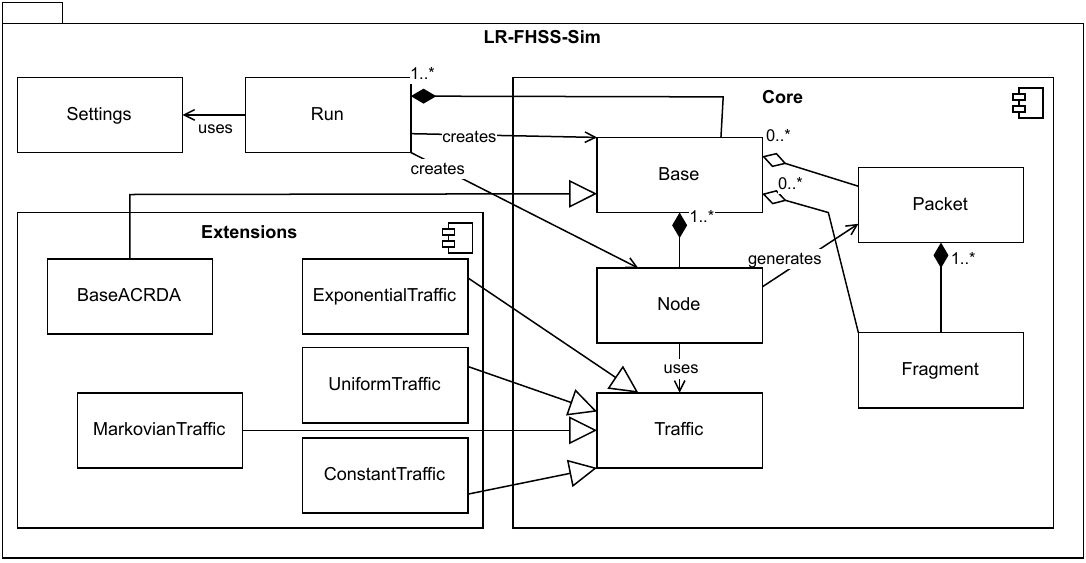}
    \caption{Simplified UML\cite{Fowler:2003} diagram of the LR-FHSS-Sim package, showing the \texttt{Core} and \texttt{Exensions} components and their respective classes.}
    \label{fig:simulator}
\end{figure*}

The LR-FHSS-Sim mainly comprises some core classes in the file \texttt{lrfhss\_core.py}. These classes represent the main structures of the simulator. They are depicted in \figurename~\ref{fig:simulator} and described below:
\begin{itemize}
    \item \textbf{Fragment}: The Fragment class is the simplest structure in the simulator. It contains its own information about duration, the Packet to which it belongs, its type (header or payload), and if it was transmitted, received with success, and, in case of collision, the information about the other collided fragments.
    \item \textbf{Packet}: The Packet class generates the packet structure with headers and payload fragments using the Fragment class. It also selects the OBW channels for each of its fragments and can deliver the next Fragment to be transmitted. The packet also contains information about the Node to which it belongs and if it was successfully received.
    \item \textbf{Node}: The Node class represents the end devices and is the one with the transmission routine. This routine runs until the end of the simulation time for each end device in the network. The transmission routine triggers all other parts of the simulation. The Node generates one new Packet to be transmitted following a specific traffic pattern, implemented by the Traffic class. Finally, it has information on the number of Packets transmitted.
    \item \textbf{Traffic}: Traffic is an abstract class used by the Node class and responsible for generating the packet inter-arrival intervals for each node following a specific traffic model. Since it is common in network simulations to evaluate different traffic models, we choose to design the traffic model as an extensible component. The traffic model class includes functions that return the interval, in seconds, to the next packet transmission. Different traffic models have been developed as extensions, as discussed later in the paper.
    \item \textbf{Base}: The Base class represents the receiver, i.e., the gateway. It contains information on the transmitted fragments and evaluates whether a fragment transmission was successful. During the configuration of the simulation, instances of the Node class are aggregated to the Base, so they can be processed during simulation. In the case of the simple Base class, it checks if there is a collision, considering that all end devices are in the Base coverage area. Moreover, it tries to decode packets when their transmission ends according to the coding rate used and stores information about how many successful packet transmissions each node made. 
\end{itemize}

There are two other files in the repository for running a simple simulation. Although they are not exactly parts of the simulation package, we provided them as examples in case someone needs a starting point when building their simulations. The first file is \texttt{settings.py}. Here, inspired by the Parametric Singleton Design Pattern~\cite{Lyon:JOT:2007}, we defined a Settings class that has all the necessary parameters to run a simulation. It makes simulations easy, as we calculate the number of headers and fragments automatically, and it presents standard inputs for all simulation parameters. Thus, one can focus only on the variables in which they are interested.

The later file is \texttt{run.py}, where we design the simulation using the network components and possibly return the numeric results we seek. 

To generate the simplest network, one needs to create the SimPy environment, the Base class, and the Nodes (end devices), add each node to the Base, add the Node transmit method to the environment process handler, and finally start the simulation for a determined amount of time. The results can be obtained by accessing the network objects of interest. All these steps and others, such as selecting different extensions, are present in the \texttt{run.py} file.

\subsection{Extensions}

An interesting feature of the simulator is the development of extensions of different signal processing and machine learning algorithms and aspects of network modeling, like traffic patterns, channel modeling, and others. Here, we present two extensions to serve as examples for future developments.

\subsubsection{Traffic Modeling} The first set of extensions refers to the transmission generation model, stored in the file \texttt{traffic.py}. Note that the traffic generator is a mandatory part of the simulation, so at least one version of this class must be available. The selection of the traffic model is made by adjusting the corresponding parameter in the \texttt{settings.py} file. For example, one implementation models the inter-arrival time as an exponentially distributed random variable, corresponding to a Poisson traffic model, possibly the most used traffic model in simulations and mathematical modeling. We have also modeled the inter-arrival time as a uniform random variable and a constant value with a small Gaussian drift to avoid collisions by the same end devices.

These three options model sparse and uniform traffic in the long term. To model less uniform traffic, we implemented a simple two-state Markov chain traffic model inspired by~\cite{Qasmi:ISWCS:2019}. We consider State $0$ as not transmitting and State $1$ as transmitting. The probability of staying in State $0$ is $p_i$, and the probability of going from State $1$ to State $0$ is $q_i$. Since this is a discrete model, we have to consider that each step represents a certain amount of time $S_M = T \pi_1$, where $T$ is the desired average inter-arrival time and $\pi_1$ is the State $1$ steady-state probability given as
\begin{align}
    \pi_1 = \frac{p_i-1}{p_i-q_i-1}.
\end{align}
Note that small values of $p_i$ imply in many transmissions (or small $T$ for a fixed $S_M$), and values of $q_i<<p_i$ represent traffic with burst behavior, where there is a higher probability of transmitting several messages subsequently.

\subsubsection{Asynchronous Contention Resolution Diversity Aloha (ACRDA)} A second extension, is the recently proposed ACRDA-enabled LR-FHSS network~\cite{Santana:IoTJ:2024}, contained in file \texttt{acrda.py}. For this extension, we need to implement a receiver with ACRDA capabilities. To do so, we extended the Base class into the BaseACRDA class and modified it accordingly. In short, we override the core Base decoding method by implementing the ACRDA decoding procedure. Moreover, we included in the BaseACRDA a memory buffer for packets and a window procedure to be added as an independent process in the SimPy environment, similar to the Node transmission. This configuration is in the \texttt{run.py} file.

\section{Results}\label{sec:results}

In this section, we present some numerical results acquired using the LR-FHSS-Sim. We start with some basic results, then compare the traffic modules presented in the previous section and show the impact when using different models. Finally, we compare different traffic models when using the ACRDA-LR-FHSS extension.
For all results, whenever not explicitly mentioned, we considered DR8 (3 headers and CR 1/3), 20 bytes of payload, simulation time of 24 hours, 100 iterations, average interval between transmissions $T$ of 900 s (4 transmissions per hour), 1 grid of 35 OBW channels, where we scale the total number of end devices multiplying by 8 (considering 8 different channel grid available), ACRDA normalized window size of 2 and normalized window step of 0.5. For the two-state Markov traffic model we used $p_i=0.99998$ and $q_i=0.15$. These two last parameters, used as an example, present a network with a burst traffic behavior, with a high probability of transmitting messages in sequence. It is important to note that we validate the simulator with literature theoretical results in~\cite{Santana:IoTJ:2024}, where the network average success behavior follows a basic mathematical framework.

\begin{figure}[tb]
    \centering
    \includegraphics[width=1\linewidth]{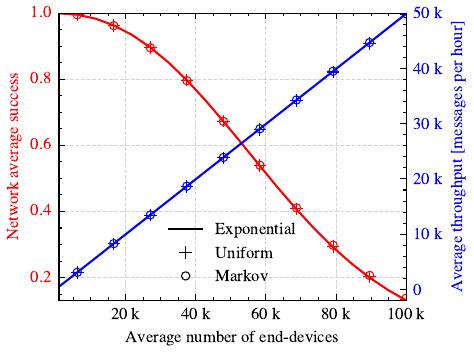}
    \caption{Network average success (red left) and throughput (blue right) for regular LR-FHSS network for different number of end devices $N$ with different traffic models.}
    \label{fig:throughput}
\end{figure}

In \figurename~\ref{fig:throughput}, we present the network's average success rate and throughput when using different device traffic models for different numbers of end devices. In blue, it is important to see that the average network throughput remains the same for all models, which means that the simulator input value of the average interval between transmissions holds true for all models. Moreover, in red, we can see that the different traffic models almost did not impact the average success of the standard LR-FHSS network. This makes sense since we are considering a long-run simulation of 24 hours with a large number of end devices. Thus, it has enough time to generate the average traffic result.

\begin{figure}[tb]
    \centering
    \includegraphics[width=1\linewidth]{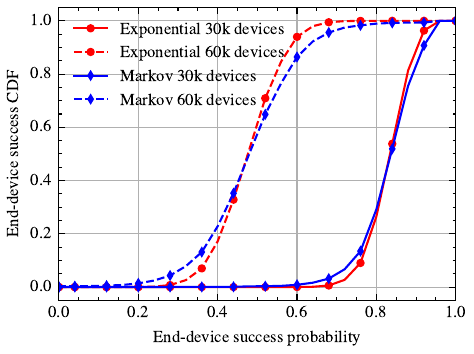}
    \caption{Cumulative distribution function of the end device success probability for different traffic models and number of end devices with 5 hours of simulation time.}
    \label{fig:cdf}
\end{figure}

\figurename~\ref{fig:cdf} depicts the cumulative distribution of the end device success probability for different traffic models when using a 5-hour simulation run. We used a reduced simulation run window to better visualize the variance caused by the Markov traffic. Here, we can see that, despite the network average success of both traffic models being the same, they present some differences. The Markov traffic model presents a slightly higher variance in the end device probability than the exponential model. This is due to the Markov traffic presenting a more unpredictable traffic generation with burst behavior. This shows us that even though the average result is similar, we still impact the performance of the end devices, as there will be more devices with lower reliability instead of a more uniform distribution. This result is relevant because we consider a case where all devices are in coverage; thus, they present the same behavior, where the traffic distribution is the only difference.

\begin{figure}[tb]
    \centering
    \includegraphics[width=1\linewidth]{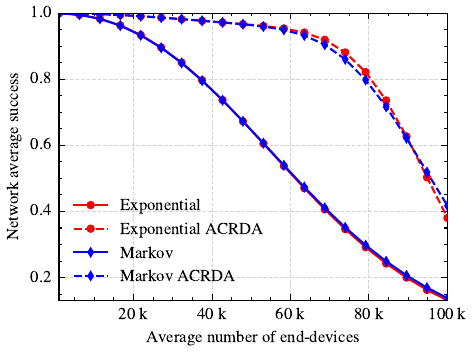}
    \caption{Network average success for regular LR-FHSS network and ACRDA-based LR-FHSS~\cite{Santana:IoTJ:2024} for different number of end devices $N$ with different traffic models.}
    \label{fig:acrda}
\end{figure}

Finally, \figurename~\ref{fig:acrda} presents the network average success of the ACRDA module~\cite{Santana:IoTJ:2024} when using the exponential and two-state Markov traffic modules. Unlike \figurename~\ref{fig:throughput}, we can see a difference in network average success between the two traffics when using the ACRDA. This result comes from the fact that devices usually tend to transmit many messages in a sequence when using Markov traffic with burst behavior. Thus, whenever two or more devices transmit simultaneously, they are expected to transmit new messages and more collisions occur. All these transmitted messages subsequently impose difficulties on the ACRDA mechanism, especially with a limited decoding sliding window, when applying the interference cancellation method and recovering the collided messages.

\section{Conclusions}\label{sec:conclusions}

In this work, using the SimPy framework, we introduced the LR-FHSS-Sim, an open-source discrete-event simulator for LR-FHSS networks in Python. We presented the simulator architecture, how it works, and how to develop new functionalities. Moreover, we presented a traffic model extension, where we can simulate different traffic patterns for the end devices and evaluate the results regarding average network success and network throughput. We showed that traffic models can present similar average performance regarding success and throughput, but there may be relevant differences in other performance statistics. 
We hope that LR-FHSS-Sim can become a useful tool for the wireless community and be further developed to include other practical extensions.

\section*{Acknowledgements}
This work or its authors have been partially supported in Finland by the Research Council of Finland (former Academy of Finland) 6G Flagship Programme (Grant Number: 346208), and by the European Union through the Interreg Aurora project ENSURE-6G (Grant Number: 20361812).; and in Brazil by CNPq (Grant Numbers: 402378/2021-0 and 305021/2021-4) and RNP/MCTIC 6G Mobile Communications Systems (Grant Number: 01245.010604/2020-14).

\bibliographystyle{IEEEtran}
\bibliography{references}

\end{document}